\documentclass[conference]{IEEEtran}
\IEEEoverridecommandlockouts
\usepackage{cite}
\usepackage{amsmath,amssymb,amsfonts}
\usepackage{algorithmic}
\usepackage{graphicx}
\usepackage{textcomp}
\usepackage{graphicx}
\usepackage{subcaption}
\usepackage{url}
\usepackage{booktabs}
\usepackage[table,xcdraw]{xcolor}
\usepackage[table,xcdraw]{xcolor}
\usepackage{xcolor}
\def\BibTeX{{\rm B\kern-.05em{\sc i\kern-.025em b}\kern-.08em
    T\kern-.1667em\lower.7ex\hbox{E}\kern-.125emX}}
\begin{document}

\title{A Sharpness Based Loss Function for Removing Out-of-Focus Blur \\
}

\author{
\IEEEauthorblockN{Uditangshu Aurangabadkar$^\star$, Darren Ramsook$^\dagger$, Anil Kokaram$^\ddagger$}
Sigmedia Group, Department of Electronic and Electrical Engineering \\
{Trinity College Dublin, Ireland}\\
$\star$aurangau@tcd.ie, $\dagger$ramsookd@tcd.ie, $\ddagger$anil.kokaram@tcd.ie \\
\textit{www.sigmedia.tv}
}

\maketitle

\begin{abstract}
The success of modern Deep Neural Network (DNN) approaches can be attributed to the use of complex optimization criteria beyond standard losses such as mean absolute error (MAE) or mean squared error (MSE). In this work, we propose a novel method of utilising a no-reference sharpness metric $Q$ introduced by Zhu and Milanfar for removing out-of-focus blur from images. We also introduce a novel dataset of real--world out-of-focus images for assessing restoration models. Our fine-tuned method produces images with a 7.5\% increase in perceptual quality (LPIPS) as compared to a standard model trained only on MAE. Furthermore, we observe a 6.7\% increase in $Q$ (reflecting sharper restorations) and 7.25\% increase in PSNR over most state-of-the-art (SOTA) methods. 
\end{abstract}


\section{Introduction}
\label{sec:introduction}

\renewcommand{\thefootnote}{\fnsymbol{footnote}}
\footnotetext[4]{This work was supported in part by a YouTube / Google Faculty award.}

The task of deblurring can be formulated as generating an image $\hat{I}$ from a blurry image $G$, such that it closely resembles the original image $I$. 
An image degraded with blur can be mathematically formulated as follows.
\begin{equation}
    G = I * H_K + \epsilon
\end{equation}
where $G$ is a degraded image, $I$ is the undegraded image, $H_K$ is a blur kernel of size $K \times K$ and  $\epsilon \in {\cal N}(0, \sigma_e^2)$ is additive noise. 

The problem can be divided into blind and non-blind depending on the prior knowledge about the parameters of the blur kernel. We can further categorize the degradation as motion blur or out-of-focus blur. Motion blur is usually caused by a sudden change in the camera position or the shutter speed. Out-of-focus blur is caused due to a change in the focal length of the camera. In real-world scenarios, a blurry image is a combination of both out-of-focus and motion blur, along with some amount of noise. In the current work, we will only deal with out-of-focus blurs.

The efficiency of DNN based architectures for restoring blurred images has given rise to several works that build on large scale architectures \cite{zamir2022restormer, kupyn2018deblurgan, 9287072}. Although these methods produce state-of-the-art (SOTA) results, their compute intensive training regimes make it difficult for deployment on resource constrained machines. This has led to the development of hybrid architectures which combine classical or iterative methods with shallow networks \cite{chen2024deep, gnanasambandam2024secrets} to produce comparable results. 
An example of this hybrid method is proposed in work \cite{lopez2023deep} by Lopez et al. where a Wiener filter is used to approximate a blur kernel, followed by a shallow network to restore artefacts generated during the process.

Although there have been works examining different architectures for the purposes of deblurring, there have been only a few that examine the role of losses in DNN architectures~\cite{tomosada2021gan}. Most architectures utilize $l_p, (p=1,2)$ norm or Welsch loss~\cite{xu2022image} as a data-fidelity term, owing to the ease of computation and differentiability. A key issue of using such losses is their inability to capture the perceptual information of an image such as sharpness, leading to additional enhancement steps such as sharpening. For instance, Deng et. al.~\cite{7340839} propose an efficient method for edge sharpening using transformed Heaviside functions.

Zhu and Milanfar~\cite{zhu2010automatic} presented a no-reference sharpness metric, $Q$ for  optimizing iterative denoising algorithms e.g. Steering Kernel Regression~\cite{takeda2007kernel} and BM3D~\cite{danielyan2011bm3d}. However, its role as a loss function in DNN models has not been examined. In this work, we propose a method of utilizing this metric as a loss in standard encoder-decoder architectures and compare it against traditional pixel losses such as MAE or MSE.  

Due to the data-intensive training regimes of DNN-based architectures, several datasets have been proposed for the tasks of motion deblurring (GoPro Dataset~\cite{Nah_2017_CVPR}), dual-pixel defocus deblurring (DPDD Dataset~\cite{abuolaim2020defocus}) and Single-Image Defocus Deblurring (RealDOF Dataset~\cite{ruan2022learning}). In addition to this, data augmentation techniques are frequently employed to produce more learning examples by synthetically degrading images with either a Gaussian or a box blur. The process of synthetic degradation, albeit effective in practice, does not capture the defocus calibration and noise profile of a consumer camera.

Inspired by the demand for high-quality training and evaluation data, we introduce a set of 305 high-resolution images, along with 3 different varying levels of blur, totalling to 1220 images for training and testing. 
The primary contributions of our paper are as follows:

\begin{itemize}
    \item A novel method of utilizing a no-reference sharpness metric as a loss function in DNN architectures for single image defocus deblurring. 
    \item A novel dataset comprising of High-Resolution images captured using a consumer camera with three different out-of-focus blur levels.
\end{itemize}

\section{Related Work}
\label{sec:related_work}
Perceptual metric based DNN models have shown to produce results comparable to SOTA in the field of image compression artefact removal, as well as image denoising~\cite{yang2018low}.

It must be noted that although employing perceptual losses for image restoration produces results closely matching the Human Visual System, most of these functions are complex and non-differentiable. Chen et al. \cite{chen2020proxiqa} model a non-differentiable loss function using a shallow Convolutional Neural Network (CNN). This is then used as a differentiable \textit{proxy} loss  in an image restoration network for removing compression artefacts. This idea was further explored by Ramsook et. al. \cite{ramsook2021differentiable} and extended to a new loss function for video, VMAF.

In this paper, we wish to use $Q$ (a no-reference metric that measures the sharpness of an image) as the loss function. A higher value corresponds to a sharper image. To measure $Q$, an image is divided into $m$ non-overlapping patches. For each patch, the matrix \textbf{$\mathbf{G}$} is created in which each row corresponds to the the horizontal and vertical gradients at each pixel site in that patch. The singular values $s_1$ and $s_2$ for a patch determine the strength of the gradients along the dominant and perpendicular directions~\cite{feng2002multiscale, zhu2009no}. These are calculated from the eigen-decomposition  of matrix $\mathbf{G^TG}$ as follows.
\begin{equation}
    \mathbf{G^TG} = \mathbf{V}
    \begin{pmatrix}
        s_1^2 & 0 \\
        0 & s_2^2
    \end{pmatrix}
    \mathbf{V^T}
    \label{eq:svd_decomp}
\end{equation}
For each patch, the local coherence $R$ is then measured as follows.
\begin{equation}
    R = \frac{s_1 - s_2}{s_1 + s_2}
    \label{eq:local_coherence}
\end{equation}
A threshold $\tau$ determines the isotropy of each patch. For anisotropic patches ($R > \tau$), i.e patches with texture, $Q$ is measured as follows.
\begin{equation}
    Q_i = s_1 \cdot \frac{s_1 - s_2}{s_1 + s_2}
    \label{eq:patch_Q}
\end{equation}
Finally, $Q$ for the entire image is calculated as the average $Q$ over all patches,
\begin{equation}
    Q = \frac{1}{N}\sum_{i=0}^{m} Q_i
    \label{eq:complete_Q}
\end{equation}
where $N$ is the total number of patches. 

The key observation to make in the presentation of $Q$ is that, unlike more complex visual quality metrics, $Q$ {\em is in fact differentiable} and can be used directly as a loss function in optimising a DNN model.

\section{Dataset}
\label{sec:dataset}

\begin{figure}
    \begin{subfigure}{0.480\linewidth}
      \includegraphics[width=\linewidth]{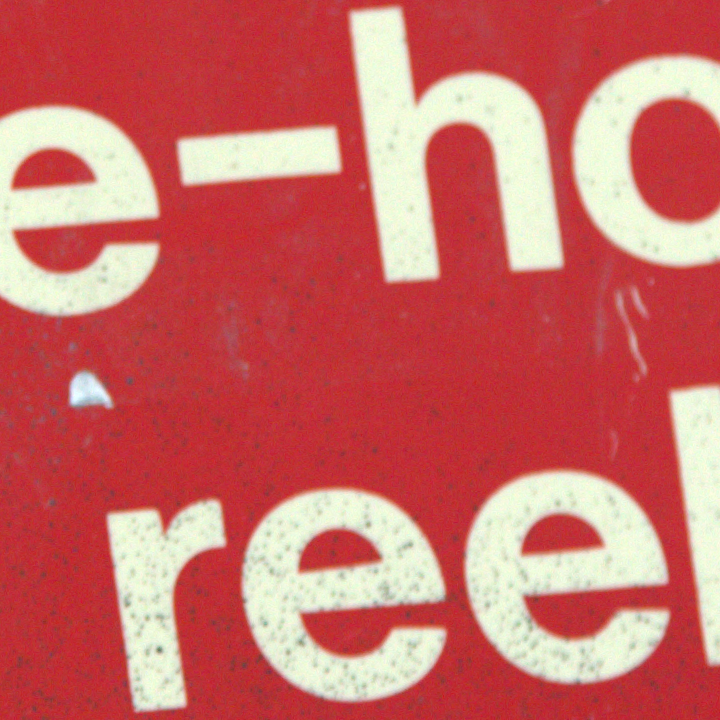}
      \caption{Original Image}
      \label{fig:original_image}
    \end{subfigure}\hfill
    \begin{subfigure}{0.480\linewidth} 
      \includegraphics[width=\linewidth]{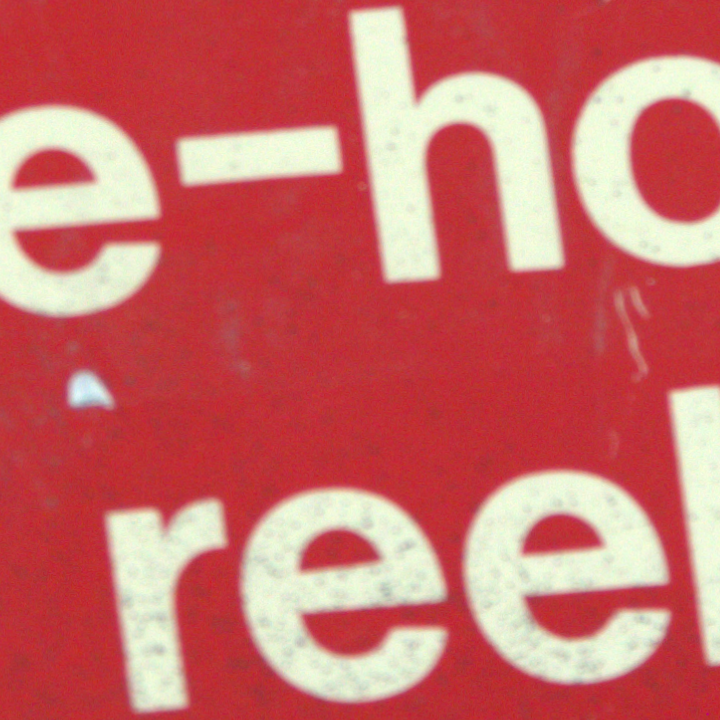}
      \caption{Low Blur Image}
      \label{fig:lowBlur_image}
    \end{subfigure}
    
    \medskip
    \begin{subfigure}{0.480\linewidth}
      \includegraphics[width=\linewidth]{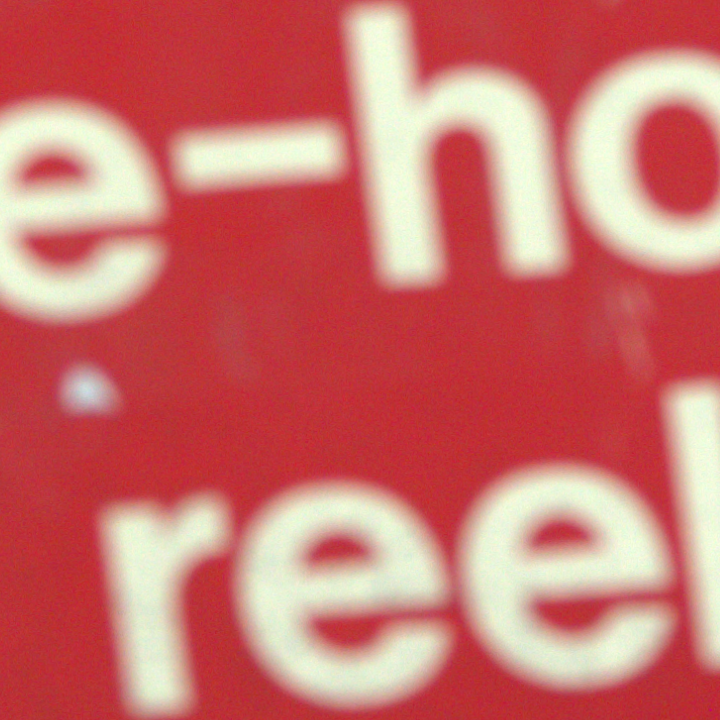}
      \caption{Medium Blur Image}
      \label{fig:medBlur_image}
    \end{subfigure}\hfill
    \begin{subfigure}{0.480\linewidth}
      \includegraphics[width=\linewidth]{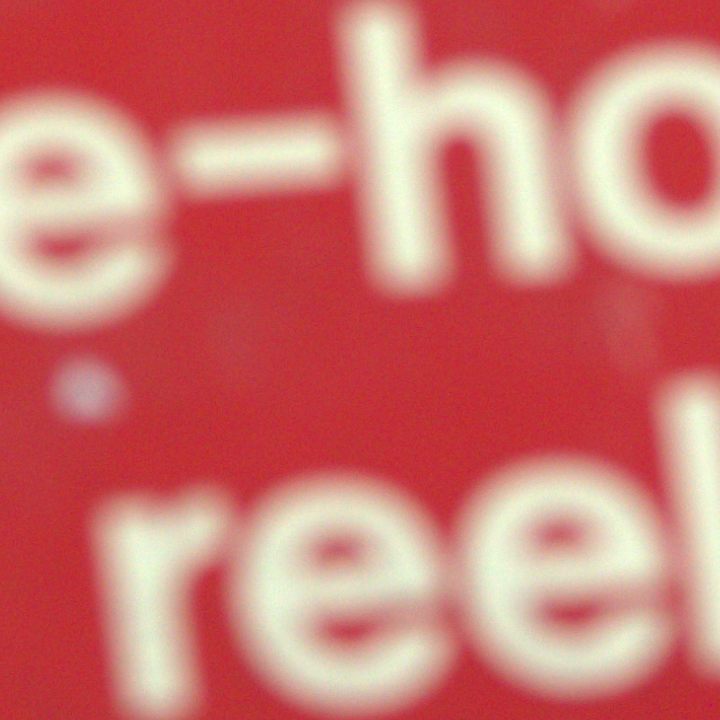}
      \caption{High Blur Image}
      \label{fig:highBlur_image}
    \end{subfigure}

    \caption{Crop of an image from the proposed dataset alongside three out-of-focus counterparts.}
    \label{fig:sigmedia_dataset}
\end{figure}

We introduce a collection of 1220 images captured using a popular consumer DSLR camera -- Canon EOS 5D Mark III ~\footnote{\url{https://www.canon.ie/for_home/product_finder/cameras/digital_slr/eos_5d_mark_iii}}. A fixed zoom and ISO were used to capture the Ground Truth (GT) image. The focus ring was divided into equal intervals of 1 millimeter or \textit{unit}. Following this, the focus ring was rotated at equal intervals of 2 units to produce 3 different out-of-focus counterparts, viz. \textit{low blur}, \textit{medium blur} and \textit{high blur}, w.r.t GT images. Each image is an 8-bit RGB image of the dimensions $5796$ $\times$ $3870$. A crop of an image from the proposed dataset can be seen in fig.~\ref{fig:sigmedia_dataset}. As the change in focus under different lighting conditions produces a color imbalance between the sharp and out-of-focus images, we utilise an automated colour transfer method introduced in work~\cite{pitie2007automated}. We also perform multi-modal image registration with 2300 iterations to align the GT and out-of-focus images. The dataset is provided in the supplementary materials~\footnote{https://github.com/aurangau/MMSP2024}. 

\section{Network Architecture}
Our model comprises of an encoder--decoder network inspired by the U-Net architecture proposed by Ronneberger et al.~\cite{ronneberger2015u}. 
A convolutional \textit{block} comprises a $3 \times 3$ convolutional layer followed by a Batch Normalization layer~\cite{ioffe2015batch} and an activation layer. We chose a differentiable non-linear activation function -- Gaussian Error Linear Unit (GELU)~\cite{hendrycks2016gaussian} as it is employed in several large scale DNN models for image generation tasks~\cite{dosovitskiy2020image}. 

\begin{figure}
    \centering    
    \includegraphics[width=0.80\linewidth]{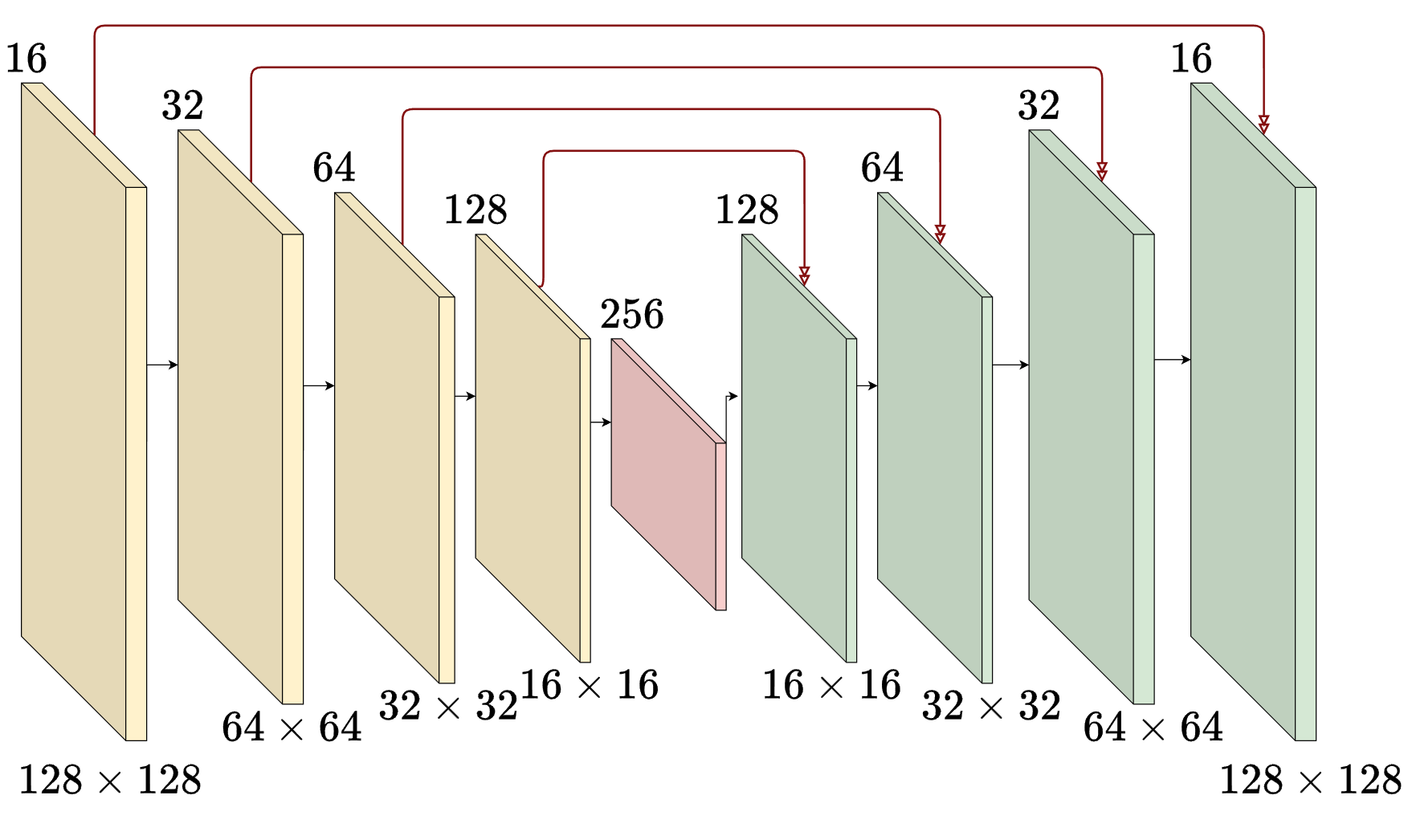}
    \caption{Proposed network architecture. Yellow, green and pink panels represent the encoder, decoder and bottleneck block. Red arrows represent the skip connections.}
    \label{fig:unet_arch}
\end{figure}

A single encoder \textit{block} comprises 5 such convolutional blocks. The encoder path comprises 4 such layers with 16, 32, 64 and 128 filters. This is followed by a bridge layer with 256 filters. A single decoder block consists of a transposed convolutional layer for upsampling, followed by a convolutional block. The decoder path comprises 4 such blocks each with 128, 64, 32 and 16 filters, respectively. The output layer is a $1 \times 1$ convolutional layer with a linear activation function. Skip connections are used to connect the encoder and decoder paths. Figure~\ref{fig:unet_arch} provides a visual representation of our model.

The network is initially trained using Mean Absolute Error (MAE) as loss as follows.
\begin{equation}
    L_1 = ||\tilde{y} - y||_1
    \label{eq:loss_mae}
\end{equation}
where $\tilde{y}$ is the predicted image and $y$ is the GT image. To fine-tune our network for producing sharper images, we introduce the sharpness metric $Q$ into the loss function.
As a higher $Q$ value corresponds to a sharper image and lower value corresponds to a blurry image, we require to minimise the negative of $Q$. Applying a coefficient $\beta$ to control the sharpness of the restored image then results in the sharpness augmented loss function $\mathcal{L}_c$ as follows.
\begin{equation}
    \mathcal{L}_c = L_1 + \beta \cdot \hat{Q}(\tilde{y})
    \label{eq:loss_comp}
\end{equation}
where $\hat{Q}(\cdot)$ = $-Q(\cdot)$.  Subsequent training steps then employ $\mathcal{L}_c$ to improve sharpness.

\section{Training}

We use 2100 images from the DIV2K~\cite{Agustsson_2017_CVPR_Workshops} and Flickr2K~\cite{Lim_2017_CVPR_Workshops} datasets for training and evaluation. 134,400 crops of size $128 \times 128$ were used for training. The degraded counterparts were generated by applying an average blur of size $5 \times 5$ to the sharp images. 

To restore real-world out-of-focus images, 77,000 crops of size 128 $\times$ 128 were used from the proposed dataset. The corresponding degraded counterparts were taken from the low blur section of our dataset. In both cases, only the Luma (Y) channel was used, as $Q$ can only be measured on that channel. 

In the case of restoring artificially degraded images, our model was trained only with $L_1$ loss for 100 epochs. To fine-tune our network, we train our model with the composite loss $\mathcal{L}_c$ detailed in the previous section for 100 epochs. During both stages of training, a constant batch size of 32 was used. A learning rate of $1e{-4}$ was used initially and was gradually reduced to $1e{-12}$. Our model consists of 4.9M parameters. The total training time was noted to be approximately 400 minutes. 

The training regime for restoring real world out-of-focus images was similar to the process detailed above. Our model was trained for 100 epochs with $l_1$ loss. The best weights were found at epoch 55. Our model was then fine-tuned using the composite loss $\mathcal{L}_c$ detailed in eq. ~\ref{eq:loss_comp} with a $\beta$ value of 0.1 for 55 epochs. The learning rate was set to $1e -4$ for the first 15 epochs, followed by a gradual decrease to $1e -12$ for the last 20 epochs. The total training time was around 392 minutes.  

The training and inference were conducted on an Intel i9-10900F@2.80 GHz CPU with a GeForce RTX2080Ti GPU. We implement the model in Tensorflow v2.0. The code for our architecture can be found in the supplementary materials~\footnote{https://github.com/aurangau/MMSP2024}. 

\section{Experimental Results}
We evaluate the performance of our proposed mechanism on synthetically blurred images, as well as, real-world out-of-focus images.

\subsection{Synthetic Dataset}
For the purposes of testing our model on artificially degraded images, we use 24 images from the Kodak dataset~\cite{KodakImages} and 500 images from the BSD500~\cite{MartinFTM01} dataset to test our model. 

\begin{figure}[]
    \begin{subfigure}{0.330\linewidth}
      \includegraphics[width=\linewidth]{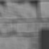}
      \caption{Blurry Image}
      \label{fig:blurry_image_comp}
    \end{subfigure}\hfill
    \begin{subfigure}{0.330\linewidth}
      \includegraphics[width=\linewidth]{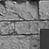}
      \caption{$\beta$ = $0$}
      \label{fig:Beta_0_restored}
    \end{subfigure}\hfill
    \begin{subfigure}{0.330\linewidth} 
      \includegraphics[width=\linewidth]{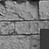}
      \caption{$\beta$ = $0.001$}
      \label{fig:Beta_001_restored}
    \end{subfigure}

    \vspace{1em}
    \begin{subfigure}{0.330\linewidth}
      \includegraphics[width=\linewidth]{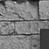}
      \caption{$\beta$ = $0.01$}
      \label{fig:Beta_001_restored}
    \end{subfigure}\hfill
    \begin{subfigure}{0.330\linewidth}
      \includegraphics[width=\linewidth]{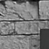}
      \caption{$\beta$ = $0.05$}
      \label{fig:Beta_005_restored}
    \end{subfigure}\hfill
    \begin{subfigure}{0.330\linewidth}
      \includegraphics[width=\linewidth]{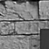}
      \caption{$\beta$ = $0.1$}
      \label{fig:Beta_0_1_restored}
    \end{subfigure}

    \caption{The visual effect of a gradual increase in $\beta$ used during fine-tuning. As $\beta$ increases, the images produced are sharper (N.B. the texture around the bricks)}
    \label{fig:beta_change}
\end{figure}

Table~\ref{tab:beta_comp} gives a comparison of using different values for the hyper-parameter $\beta$ in equation~(\ref{eq:loss_comp}) to generate images with varying degrees of sharpness and quality for the Kodak dataset. It must be noted that although assigning a larger weight to $Q$ generates sharper images, it does not necessarily produce images with a higher PSNR. Assigning a smaller weight to $Q$, however, produces images with a higher PSNR and SSIM than using $l_1$ norm alone (highlighted in gray). An example of this effect can be seen for a crop from figure \textit{Kodim01} (fig.~\ref{fig:beta_change}).

\begin{table}[]
\centering
\begin{tabular}{ccccc}
\toprule
\textbf{$\beta$}  & \textbf{PSNR (dB)} & \textbf{SSIM} & \textbf{$Q$} & \textbf{LPIPS} \\ \midrule
\rowcolor[HTML]{C0C0C0} 
\textbf{0}     & 35.069            & 0.944        & 0.153     & 0.127         \\
\textbf{0.001} & 35.102            & 0.945        & 0.152     & 0.117         \\
\textbf{0.01}  & 35.101            & 0.945        & 0.154     & 0.117         \\
\textbf{0.05}  & 34.844            & 0.945        & 0.166     & 0.122         \\
\textbf{0.1}   & 33.641            & 0.940        & 0.183     & 0.127         \\ \bottomrule
\end{tabular}
\caption{Comparison of weighting coefficient $\beta$ for fine-tuning our model. An increase in $\beta$ leads to an increase in sharpness ($Q$). }
\label{tab:beta_comp}
\end{table}

A $\beta$ value of $0.01$ produces an image with a higher PSNR and a higher $Q$ than the baseline model trained only using $L_1$ loss. We also discover that using a $\beta$ value higher than $0.5$ produces images with severe ringing artefacts (see supplementary materials). To understand the impact of $\beta$ on the perceptual quality of the restored images, we also measure LPIPS~\cite{zhang2018unreasonable}. A lower LPIPS corresponds to higher perceptual quality. 

We compare our model against DNN based architectures and iterative algorithms. For DNN based architectures, we select a lightweight deblurring model titled DDNet~\cite{lopez2023deep} and a SOTA deblurring model originally developed for motion deblurring titled XY--Deblur~\cite{ji2022xydeblur}. The model was re-trained on our training data and was found to be effective at restoring out-of-focus images. For iterative algorithms, we select some of the methods introduced in work~\cite{belyaev2022black}, namely the accelerated Landweber (aL) like iterations, modified Richardson-Lucy (mRL) method and modified Levenberg-Marquardt (mLM) method. For all three methods, the maximum number of iterations was set to 250. We also select two iterative methods that are based on regularization by denoising (RED) titled Fixed-Point (FP) and Steepest Descent Method (SDM) introduced by Romano, Elad and Milanfar~\cite{romano2017little}. We utilize PSNR, SSIM, $Q$ and LPIPS to measure the efficacy of the aforementioned restoration methods over the Kodak and BSD500 datasets, as well as, measuring the inference time (in seconds). Table~\ref{tab:comparison_table_artificial} provides a metric comparison averaged over the two datasets. 

\begin{table}[]
\begin{tabular}{@{}lccccc@{}}
\toprule
\textbf{Method Name}            & \multicolumn{1}{l}{\textbf{PSNR (dB)}} & \multicolumn{1}{l}{\textbf{SSIM}} & \multicolumn{1}{l}{\textbf{$Q$}} & \multicolumn{1}{l}{\textbf{LPIPS}} & \multicolumn{1}{l}{\textbf{Time (s)}} \\ \midrule
DDNet~\cite{lopez2023deep}                           & 29.559                                 & 0.842                             & 0.130                          & 0.271                              & 0.019                                      \\
XY--Deblur~\cite{ji2022xydeblur}                             & \underline{38.288}                                 & 0.970                             & 0.147                          & 0.071                              & 0.019                                     \\
aL~\cite{belyaev2022black}                             & 29.362                                 & 0.882                             & 0.136                          & 0.229                              & 0.009                                     \\
mLM                             & 37.282                                 & \underline{0.971}                             & 0.135                          & \underline{0.059}                              & 0.003                                     \\
mRL                             & 29.658                                 & 0.891                             & 0.134                          & 0.212                              & 0.012                                      \\
RED-FP~\cite{romano2017little}                          & 33.137                                 & 0.911                             & \underline{0.154}                          & 0.181                              & 47.800                                     \\
RED-SDM                         & 32.835                                 & 0.909                             & 0.151                          & 0.181                              & 328.259                                     \\ \midrule
\textit{\textbf{Ours (w/o. FT)}} & \textit{\textbf{34.875}}               & \textit{\textbf{0.946}}           & \textit{\textbf{0.149}}        & \textit{\textbf{0.116}}            & \textit{\textbf{0.052}}                   \\
\textit{\textbf{Ours (w. FT)}}   & \textit{\textbf{34.890}}               & \textit{\textbf{0.947}}           & \textit{\textbf{0.150}}        & \textit{\textbf{0.107}}            & \textit{\textbf{0.063}}                   \\ \bottomrule
\end{tabular}
\caption{Comparison of restoration techniques averaged over Kodak24 and BSD500 datasets. Our method produces images with higher perceptual quality (SSIM, LPIPS) and sharpness ($Q$).}
\label{tab:comparison_table_artificial}
\end{table}

\begin{figure}
    \centering
    
    \subfloat[Original]{\includegraphics[width=0.16\textwidth]{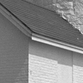}} \hfill
    \subfloat[Blurry]{\includegraphics[width=0.16\textwidth]{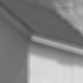}} \hfill
    \subfloat[DDNet\cite{lopez2023deep} ]{\includegraphics[width=0.16\textwidth]{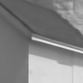}} \\
    
    \subfloat[XY--Deblur \cite{ji2022xydeblur}]{\includegraphics[width=0.16\textwidth]{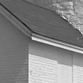}} \hfill
    \subfloat[aL \cite{belyaev2022black}]{\includegraphics[width=0.16\textwidth]{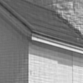}} \hfill
    \subfloat[mLM]{\includegraphics[width=0.16\textwidth]{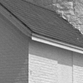}} \\
    
    \subfloat[RED-SDM]{\includegraphics[width=0.16\textwidth]{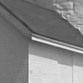}} \hfill
    \subfloat[Ours (w/o. FT)]{\includegraphics[width=0.16\textwidth]{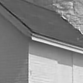}} \hfill
    \subfloat[Ours (w. FT)]{\includegraphics[width=0.16\textwidth]{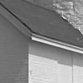}}
    
    \caption{Comparison of methods for removing synthetic out-of-focus blur. Our fine-tuned method produces a sharper edge around the roof.}
    \label{fig:all_comp}
\end{figure}

A crop from the image \textit{Kodim19} can be seen in fig.~\ref{fig:all_comp}. The fine-tuned model produces sharper images with a higher $Q$. It must be noted that although iterative methods such as RED-SDM and RED-FP produce images with a slightly higher $Q$, metrics such as LPIPS, PSNR and SSIM of the method are relatively low (table~\ref{tab:comparison_table_artificial}). Other iterative algorithms such as aL and mLM produce images with diagonal stripes and checkerboard artefacts. DDNet fails to recover the textural details of the image. Even though XY--Deblur recovers these details, the sharpness of the image is still lower than our methods. 

\begin{figure}
    \centering
    \includegraphics[width=\linewidth]{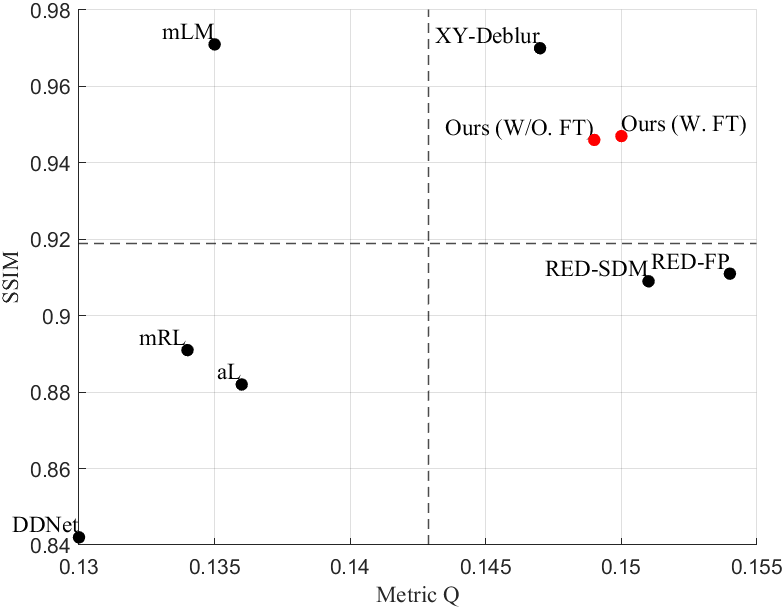}
    \caption{Comparison between sharpness ($Q$) and perceptual quality (SSIM). Our methods (highlighted in red) generate images with higher structural similarity, as well as sharpness.}
    \label{fig:QVSSIM_pllot}
\end{figure}

It must be noted that although our technique does not outperform other SOTA methods in terms of a single metric (underlined in table~\ref{tab:comparison_table_artificial}), we achieve an overall balanced performance as evident from the plot in fig.~\ref{fig:QVSSIM_pllot}. Our methods (highlighted in red) produce images with high perceptual quality (SSIM), as well as, being sharper than most restoration methods. The vertical and horizontal dashed lines in black denote the average values for $Q$ and SSIM respectively for the restoration models mentioned in table~\ref{tab:comparison_table_artificial}.

\subsection{Real Dataset}
For testing our model on real-world out-of-focus images, we use 188 crops of size $128 \times 128$ from the low blur section of our proposed dataset. We note that in order to restore images from the medium and high blur sections, a much more complex and sophisticated architecture must be devised. 

\begin{figure}
    \begin{subfigure}{0.495\linewidth}
      \includegraphics[width=\linewidth]{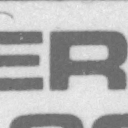}
      \caption{Original Image}
      \label{fig:original_compReal_image}
    \end{subfigure}\hfill
    \begin{subfigure}{0.495\linewidth} 
      \includegraphics[width=\linewidth]{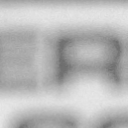}
      \caption{Out-of-focus image}
      \label{fig:lowBlur_compReal_image}
    \end{subfigure}
    
    \medskip
    \begin{subfigure}{0.495\linewidth}
      \includegraphics[width=\linewidth]{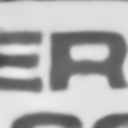}
      \caption{Image w/o. Fine-tuning}
      \label{fig:noFT_compReal_image}
    \end{subfigure}\hfill
    \begin{subfigure}{0.495\linewidth}
      \includegraphics[width=\linewidth]{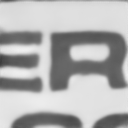}
      \caption{Image w. Fine-tuning}
      \label{fig:FT_compReal_image}
    \end{subfigure}

    \caption{Visual comparison of image restored with and w/o using $Q$ for fine-tuning. The fine-tuned method produces sharper edges.}
    \label{fig:real_image_comp}
\end{figure}

Figure~\ref{fig:real_image_comp} demonstrates the effect of fine--tuning our model with $\beta$ value of $0.1$. Although this produces images with a lower PSNR, metrics such as LPIPS and $Q$ are significantly higher. This can be attested by the sharper edges in the generated image (fig. ~\ref{fig:FT_compReal_image}).

\begin{figure}
    \begin{subfigure}{0.330\linewidth}
      \includegraphics[width=\linewidth]{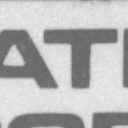}
      \caption{Original Image}
      \label{fig:original_real_image}
    \end{subfigure}\hfill
    \begin{subfigure}{0.330\linewidth}
      \includegraphics[width=\linewidth]{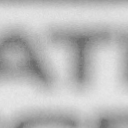}
      \caption{Blurry Image}
      \label{fig:blurry_real_image}
    \end{subfigure}\hfill
    \begin{subfigure}{0.330\linewidth} 
      \includegraphics[width=\linewidth]{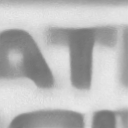}
  \caption{Restormer~\cite{zamir2022restormer}}
      \label{fig:restormer_real_image}
    \end{subfigure}

    \vspace{1em}
    \begin{subfigure}{0.330\linewidth}
      \includegraphics[width=\linewidth]{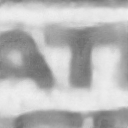}
      \caption{IFAN~\cite{lee2021iterative}}
      \label{fig:ifan_real_image}
    \end{subfigure}\hfill
    \begin{subfigure}{0.330\linewidth}
      \includegraphics[width=\linewidth]{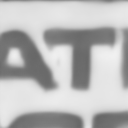}
      \caption{XY--Deblur~\cite{ji2022xydeblur}}
      \label{fig:xyd_real_image}
    \end{subfigure}\hfill
    \begin{subfigure}{0.330\linewidth}
      \includegraphics[width=\linewidth]{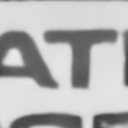}
      \caption{Ours (FT)}
      \label{fig:Qsharp_real_image}
    \end{subfigure}

    \caption{Visual Comparison of DNN-based restoration artefacts for real-world out-of-focus images. Our method produces the least amount of restoration artefacts.}
    \label{fig:real_comparisons}
\end{figure}

We compare our model against three leading DNN based restoration methods -- XY--Deblur, Restormer~\cite{zamir2022restormer} and IFAN~\cite{lee2021iterative}. A visual comparison can be seen in figure~\ref{fig:real_comparisons}. Methods such as IFAN and Restormer do not produce sharp images. As compared to XY--Deblur, our fine-tuned method produces sharper images without restoration artefacts. A quantitative comparison of the methods is given in table~\ref{tab:real_dataset_comp}.

\begin{table}[]
\begin{tabular}{@{}lccccc@{}}
\toprule
\textbf{Algorithm}              & \textbf{PSNR (dB)}       & \textbf{SSIM}           & \textbf{$Q$}              & \textbf{LPIPS}          & \textbf{Time (s)}   \\ \midrule
XY--Deblur~\cite{ji2022xydeblur}      & 28.131 & 0.784 & 0.082 & 0.668 & 0.012 \\
IFAN~\cite{lee2021iterative}      & 23.728 & 0.668 & 0.044 & \underline{0.408} & 0.022 \\
Restormer~\cite{zamir2022restormer} & 22.594 & 0.650 & 0.042 & 0.418 & 0.066 \\
\textit{\textbf{Ours (w/o FT)}} & \underline{\textit{\textbf{28.788}}} & \textit{\textbf{0.792}} & \textit{\textbf{0.079}} & \textit{\textbf{0.450}} & \textit{\textbf{0.040}} \\
\textit{\textbf{Ours (w. FT)}}  & \textit{\textbf{28.728}} & \underline{\textit{\textbf{0.792}}} & \underline{\textit{\textbf{0.084}}} & \textit{\textbf{0.447}} & \textit{\textbf{0.038}} \\ \bottomrule
\end{tabular}
\caption{Comparison  of DNN based restoration techniques on real-world out-of-focus images.}
\label{tab:real_dataset_comp}
\end{table}

\section{Conclusion}
We have proposed a method of utilising a no-reference sharpness metric $Q$ as a loss function in a DNN-based model for removing out-of-focus blur. We have shown that such an approach produces sharper images with an approximate increase of $7\%$ in perceptual quality score (LPIPS) as compared to a traditional loss such as MAE. Furthermore, we observe an average increase of approximately $6.7\%$ in $Q$ and $7.25\%$ in PSNR over most SOTA restoration models. 

We also introduce a novel dataset comprising of three different levels of blur and their corresponding sharp counterparts. We show that our method is also effective at restoring real world out-of-focus images and note an average increase of $63\%$ in $Q$ and $16\%$ in PSNR over DNN-based restoration models.  

For future work, we plan on comparing our proposed loss function with other popular complex perceptual losses (LPIPS). We also plan on utilising the proposed loss function to fine-tune SOTA DNN models such as XY--Deblur and Restormer. 

\bibliographystyle{IEEEbib}
\bibliography{refs}

\begin{thebibliography}{10}

\bibitem{zamir2022restormer}
Syed~Waqas Zamir, Aditya Arora, Salman Khan, Munawar Hayat, Fahad~Shahbaz Khan, and Ming-Hsuan Yang,
\newblock ``Restormer: Efficient transformer for high-resolution image restoration,''
\newblock in {\em Proceedings of the IEEE/CVF conference on computer vision and pattern recognition}, 2022, pp. 5728--5739.

\bibitem{kupyn2018deblurgan}
Orest Kupyn, Volodymyr Budzan, Mykola Mykhailych, Dmytro Mishkin, and Ji{\v{r}}{\'\i} Matas,
\newblock ``Deblurgan: Blind motion deblurring using conditional adversarial networks,''
\newblock in {\em Proceedings of the IEEE conference on computer vision and pattern recognition}, 2018, pp. 8183--8192.

\bibitem{9287072}
Zeyu Jiang, Xun Xu, Chao Zhang, and Ce~Zhu,
\newblock ``Multianet: a multi-attention network for defocus blur detection,''
\newblock in {\em 2020 IEEE 22nd International Workshop on Multimedia Signal Processing (MMSP)}, 2020, pp. 1--6.

\bibitem{chen2024deep}
Liang Chen, Jiawei Zhang, Zhenhua Li, Yunxuan Wei, Faming Fang, Jimmy Ren, and Jinshan Pan,
\newblock ``Deep richardson--lucy deconvolution for low-light image deblurring,''
\newblock {\em International Journal of Computer Vision}, vol. 132, no. 2, pp. 428--445, 2024.

\bibitem{gnanasambandam2024secrets}
Abhiram Gnanasambandam, Yash Sanghvi, and Stanley~H Chan,
\newblock ``The secrets of non-blind poisson deconvolution,''
\newblock {\em IEEE Transactions on Computational Imaging}, 2024.

\bibitem{lopez2023deep}
Santiago L{\'o}pez-Tapia, Javier Mateos, Rafael Molina, and Aggelos~K Katsaggelos,
\newblock ``Deep robust image restoration using the moore-penrose blur inverse,''
\newblock in {\em 2023 IEEE International Conference on Image Processing (ICIP)}. IEEE, 2023, pp. 775--779.

\bibitem{tomosada2021gan}
Hiroki Tomosada, Takahiro Kudo, Takanori Fujisawa, and Masaaki Ikehara,
\newblock ``Gan-based image deblurring using dct loss with customized datasets,''
\newblock {\em IEEE Access}, vol. 9, pp. 135224--135233, 2021.

\bibitem{xu2022image}
Zhenhua Xu, Jiancheng Lai, Jun Zhou, Huasong Chen, Hongkun Huang, and Zhenhua Li,
\newblock ``Image deblurring using a robust loss function,''
\newblock {\em Circuits, Systems, and Signal Processing}, pp. 1--31, 2022.

\bibitem{7340839}
Liang-Jian Deng, Weihong Guo, Ting-Zhu Huang, and Xi-Le Zhao,
\newblock ``Heaviside image edge sharpening,''
\newblock in {\em 2015 IEEE 17th International Workshop on Multimedia Signal Processing (MMSP)}, 2015, pp. 1--5.

\bibitem{zhu2010automatic}
Xiang Zhu and Peyman Milanfar,
\newblock ``Automatic parameter selection for denoising algorithms using a no-reference measure of image content,''
\newblock {\em IEEE transactions on image processing}, vol. 19, no. 12, pp. 3116--3132, 2010.

\bibitem{takeda2007kernel}
Hiroyuki Takeda, Sina Farsiu, and Peyman Milanfar,
\newblock ``Kernel regression for image processing and reconstruction,''
\newblock {\em IEEE Transactions on image processing}, vol. 16, no. 2, pp. 349--366, 2007.

\bibitem{danielyan2011bm3d}
Aram Danielyan, Vladimir Katkovnik, and Karen Egiazarian,
\newblock ``Bm3d frames and variational image deblurring,''
\newblock {\em IEEE Transactions on image processing}, vol. 21, no. 4, pp. 1715--1728, 2011.

\bibitem{Nah_2017_CVPR}
Seungjun Nah, Tae~Hyun Kim, and Kyoung~Mu Lee,
\newblock ``Deep multi-scale convolutional neural network for dynamic scene deblurring,''
\newblock in {\em CVPR}, July 2017.

\bibitem{abuolaim2020defocus}
Abdullah Abuolaim and Michael~S Brown,
\newblock ``Defocus deblurring using dual-pixel data,''
\newblock in {\em European Conference on Computer Vision}. Springer, 2020, pp. 111--126.

\bibitem{ruan2022learning}
Lingyan Ruan, Bin Chen, Jizhou Li, and Miuling Lam,
\newblock ``Learning to deblur using light field generated and real defocus images,''
\newblock in {\em Proceedings of the IEEE/CVF Conference on Computer Vision and Pattern Recognition}, 2022, pp. 16304--16313.

\bibitem{yang2018low}
Qingsong Yang, Pingkun Yan, Yanbo Zhang, Hengyong Yu, Yongyi Shi, Xuanqin Mou, Mannudeep~K Kalra, Yi~Zhang, Ling Sun, and Ge~Wang,
\newblock ``Low-dose ct image denoising using a generative adversarial network with wasserstein distance and perceptual loss,''
\newblock {\em IEEE transactions on medical imaging}, vol. 37, no. 6, pp. 1348--1357, 2018.

\bibitem{chen2020proxiqa}
Li-Heng Chen, Christos~G Bampis, Zhi Li, Andrey Norkin, and Alan~C Bovik,
\newblock ``Proxiqa: A proxy approach to perceptual optimization of learned image compression,''
\newblock {\em IEEE Transactions on Image Processing}, vol. 30, pp. 360--373, 2020.

\bibitem{ramsook2021differentiable}
Darren Ramsook, Anil Kokaram, Noel O'Connor, Neil Birkbeck, Yeping Su, and Balu Adsumilli,
\newblock ``A differentiable vmaf proxy as a loss function for video noise reduction,''
\newblock in {\em Applications of Digital Image Processing XLIV}. SPIE, 2021, vol. 11842, pp. 290--301.

\bibitem{feng2002multiscale}
XiaoGuang Feng and Peyman Milanfar,
\newblock ``Multiscale principal components analysis for image local orientation estimation,''
\newblock in {\em Conference Record of the Thirty-Sixth Asilomar Conference on Signals, Systems and Computers, 2002.} IEEE, 2002, vol.~1, pp. 478--482.

\bibitem{zhu2009no}
Xiang Zhu and Peyman Milanfar,
\newblock ``A no-reference sharpness metric sensitive to blur and noise,''
\newblock in {\em 2009 international workshop on quality of multimedia experience}. IEEE, 2009, pp. 64--69.

\bibitem{pitie2007automated}
Fran{\c{c}}ois Piti{\'e}, Anil~C Kokaram, and Rozenn Dahyot,
\newblock ``Automated colour grading using colour distribution transfer,''
\newblock {\em Computer Vision and Image Understanding}, vol. 107, no. 1-2, pp. 123--137, 2007.

\bibitem{ronneberger2015u}
Olaf Ronneberger, Philipp Fischer, and Thomas Brox,
\newblock ``U-net: Convolutional networks for biomedical image segmentation,''
\newblock in {\em Medical image computing and computer-assisted intervention--MICCAI 2015: 18th international conference, Munich, Germany, October 5-9, 2015, proceedings, part III 18}. Springer, 2015, pp. 234--241.

\bibitem{ioffe2015batch}
Sergey Ioffe and Christian Szegedy,
\newblock ``Batch normalization: Accelerating deep network training by reducing internal covariate shift,''
\newblock in {\em International conference on machine learning}. pmlr, 2015, pp. 448--456.

\bibitem{hendrycks2016gaussian}
Dan Hendrycks and Kevin Gimpel,
\newblock ``Gaussian error linear units (gelus),''
\newblock {\em arXiv preprint arXiv:1606.08415}, 2016.

\bibitem{dosovitskiy2020image}
Alexey Dosovitskiy, Lucas Beyer, Alexander Kolesnikov, Dirk Weissenborn, Xiaohua Zhai, Thomas Unterthiner, Mostafa Dehghani, Matthias Minderer, Georg Heigold, Sylvain Gelly, et~al.,
\newblock ``An image is worth 16x16 words: Transformers for image recognition at scale,''
\newblock {\em arXiv preprint arXiv:2010.11929}, 2020.

\bibitem{Agustsson_2017_CVPR_Workshops}
Eirikur Agustsson and Radu Timofte,
\newblock ``Ntire 2017 challenge on single image super-resolution: Dataset and study,''
\newblock in {\em The IEEE Conference on Computer Vision and Pattern Recognition (CVPR) Workshops}, July 2017.

\bibitem{Lim_2017_CVPR_Workshops}
Bee Lim, Sanghyun Son, Heewon Kim, Seungjun Nah, and Kyoung~Mu Lee,
\newblock ``Enhanced deep residual networks for single image super-resolution,''
\newblock in {\em The IEEE Conference on Computer Vision and Pattern Recognition (CVPR) Workshops}, July 2017.

\bibitem{KodakImages}
``Kodak lossless true color image suite. [online]. available: http://r0k.us/graphics/kodak/,'' 2007.

\bibitem{MartinFTM01}
D.~Martin, C.~Fowlkes, D.~Tal, and J.~Malik,
\newblock ``A database of human segmented natural images and its application to evaluating segmentation algorithms and measuring ecological statistics,''
\newblock in {\em Proc. 8th Int'l Conf. Computer Vision}, July 2001, vol.~2, pp. 416--423.

\bibitem{zhang2018unreasonable}
Richard Zhang, Phillip Isola, Alexei~A Efros, Eli Shechtman, and Oliver Wang,
\newblock ``The unreasonable effectiveness of deep features as a perceptual metric,''
\newblock in {\em Proceedings of the IEEE conference on computer vision and pattern recognition}, 2018, pp. 586--595.

\bibitem{ji2022xydeblur}
Seo-Won Ji, Jeongmin Lee, Seung-Wook Kim, Jun-Pyo Hong, Seung-Jin Baek, Seung-Won Jung, and Sung-Jea Ko,
\newblock ``Xydeblur: Divide and conquer for single image deblurring,''
\newblock in {\em Proceedings of the IEEE/CVF conference on computer vision and pattern recognition}, 2022, pp. 17421--17430.

\bibitem{belyaev2022black}
Alexander~G Belyaev and Pierre-Alain Fayolle,
\newblock ``Black-box image deblurring and defiltering,''
\newblock {\em Signal Processing: Image Communication}, vol. 108, pp. 116833, 2022.

\bibitem{romano2017little}
Yaniv Romano, Michael Elad, and Peyman Milanfar,
\newblock ``The little engine that could: Regularization by denoising (red),''
\newblock {\em SIAM Journal on Imaging Sciences}, vol. 10, no. 4, pp. 1804--1844, 2017.

\bibitem{lee2021iterative}
Junyong Lee, Hyeongseok Son, Jaesung Rim, Sunghyun Cho, and Seungyong Lee,
\newblock ``Iterative filter adaptive network for single image defocus deblurring,''
\newblock in {\em Proceedings of the IEEE/CVF conference on computer vision and pattern recognition}, 2021, pp. 2034--2042.

\end{thebibliography}

\end{document}